\documentclass[%
 reprint,
 amsmath,amssymb,
 aps,
]{revtex4-2}

\usepackage{graphicx}
\usepackage{dcolumn}
\usepackage{bm}
\usepackage{orcidlink}
\usepackage{threeparttable}
\usepackage{booktabs}

\begin{document}

\preprint{APS/123-QED}

\title{Neutrino production at the extreme vicinity of supermassive black hole in NGC 1068}

\author{Zhi-Peng Ma\,\orcidlink{0009-0007-0717-3667}}
 \affiliation{Department of Astronomy, School of Physics, Huazhong University of Science and Technology, Wuhan, Hubei 430074, China}
\author{Kai Wang\,\orcidlink{0000-0003-4976-4098}}%
 \email{kaiwang@hust.edu.cn}
\affiliation{%
 Department of Astronomy, School of Physics, Huazhong University of Science and Technology, Wuhan, Hubei 430074, China}%

\date{\today}

\begin{abstract}
The identification of NGC~1068 as an extragalactic high-energy neutrino source raises the question of the physical conditions required for neutrino production. Motivated by the newly updated IceCube data (down to $\sim 0.2 \,\rm TeV$) for NGC 1068, we present a statistical, model-independent inference primarily constrained by the multi-messenger observations. We calculate neutrino production from hadronic channels (including hadronuclear and photomeson processes) and the accompanying cascade gamma rays, and then employ the Latin hypercube sampling (LHS) method to explore the parameter space comprehensively, constrained by observational data. According to our results, the emission region is tightly limited: $R \gtrsim 10\,R_{\rm s}$ is excluded at $\sim 5\sigma$, and $R \gtrsim 6\,R_{\rm s}$ at $\sim 3\sigma$, where $R_{\rm s}$ is Schwarzschild radius of supermassive black hole (SMBH). Furthermore, reproducing $\gtrsim 80\%$ of the observed neutrino flux further restricts the emission to $R \lesssim 3R_{\rm s}$. Finally, we explore the possible particle acceleration mechanisms and find that stochastic acceleration naturally satisfies all constraints, whereas alternative mechanisms require extreme, fine-tuning parameters. We therefore identify stochastic acceleration as the most viable origin of the neutrino emission from NGC~1068.
\end{abstract}

\maketitle

\section{Introduction}
\label{sec:intro}
In 2022, the IceCube Collaboration reported compelling evidence for 1--10~TeV neutrino emission from the Seyfert~II galaxy NGC~1068 at a significance of $4.2\sigma$~\cite{icecube2022evidence}, establishing it as the first identified extragalactic persistent source of high-energy neutrinos. Subsequent searches for X-ray-bright AGN have revealed suggestive excesses in other Seyfert galaxies~\cite{abbasi2026evidence,sommani2025two,neronov2024neutrino}, whereas NGC~1068 remains the most significant source. Recent IceCube observations provide further motivation to investigate the physical origin of the neutrino emission from NGC 1068. The latest analysis reports a \(3.5\sigma\) excess from NGC 1068 and finds that its measured flux is consistent with expectations for steady emission~\citep{abbasi2025all}. These results strengthen the case for NGC 1068 as a persistent neutrino source and motivate further investigation of the underlying particle-acceleration and neutrino-production mechanisms.

A key feature of this source is its relatively low MeV--GeV gamma-ray flux~\cite{murase2022hidden,das2024revealing}, which is difficult to reconcile with the observed neutrino emission. This tension suggests efficient gamma-ray attenuation, implying that hadronic interactions occur in a dense radiation environment where gamma rays are suppressed via pair production while neutrinos escape~\cite{murase2016hidden,murase2020hidden}. In the standard disk--corona scenario, such conditions can naturally arise in the corona, a compact region populated by X-ray photons produced via Comptonization of disk photons by energetic electrons~\cite{cackett2021reverberation,shapiro1976two,zdziarski1996broad,poutanen1996two,haardt1997x}. In this region, nonthermal protons produce neutrinos via $p\gamma$ interactions with radiation fields or via $pp$ interactions with ambient matter~\cite{murase2022hidden,inoue2022high,ma2025seyfert}. The accompanying gamma rays initiate electromagnetic cascades through pair production, redistributing their energy to sub-GeV bands and alleviating the gamma-ray--neutrino tension (e.g., Refs.~\cite{murase2022hidden,das2024revealing}).

Within this framework, several mechanisms for nonthermal proton acceleration have been proposed, including diffusive shock acceleration (DSA) at accretion shocks or in AGN-driven winds, as well as magnetic dissipation processes such as stochastic acceleration (SA) in turbulence and acceleration in magnetic reconnection (MR) regions (e.g., Refs.~\cite{inoue2020origin,blandford1987particle,bell2013cosmic,malkov2001nonlinear,inoue2022high,ma2025seyfert,dermer1995stochastic,kheirandish2021high,murase2020hidden,lemoine2025neutrinos,mbarek2024interplay,fiorillo2024tev,fiorillo2025contribution,saurenhaus2026constraining,karavola2025neutrino,karavola2026diffuse,passos2026high}). These mechanisms can broadly reproduce the observed neutrino flux. Previous studies have derived multimessenger constraints on the energetics and source size, typically favoring compact regions with $R \lesssim 15 R_{\rm s}$ (low-$\beta$ plasma) or $R \lesssim 3 R_{\rm s}$ (high-$\beta$ plasma)~\cite{das2024revealing,das2026multimessenger}. More recent work within stochastic acceleration scenarios suggests $R \lesssim 10 R_{\rm s}$~\cite{carpio2026multimessenger,testagrossa2026neutrinos}. However, previous studies are largely model-dependent and lack both a statistical assessment of the constraints and a systematic exploration of the full parameter space.

More importantly, very recent observations introduce new challenges. Ref.~\cite{abbasi2026evidence} reports a neutrino spectrum consistent with an unbroken power law with spectral index $3.4\pm0.2$, extending over 0.2--20.6~TeV at 95\% C.L. The inferred flux at 0.2~TeV exceeds the gamma-ray flux by at least two orders of magnitude. This exacerbates the gamma-ray--neutrino tension and poses an additional difficulty: reproducing a broad, well-defined power-law spectrum over nearly two orders of magnitude in energy within a single framework. These stringent observations provide an opportunity to probe the physical conditions of the emission region.

In this work, based on the newly released IceCube data for NGC 1068, we present a model-independent approach to constrain the physical conditions for neutrino production in the comprehensive parameter space. Under minimal physical assumptions, we develop a spectral-shape-driven method that directly exploits the observed neutrino and gamma-ray spectra to map the allowed parameter space. This approach enables us to identify the regions favored and excluded and to infer implications for the underlying particle acceleration mechanisms.

\begin{figure*}[t]  
    \centering
    \includegraphics[width=\textwidth]{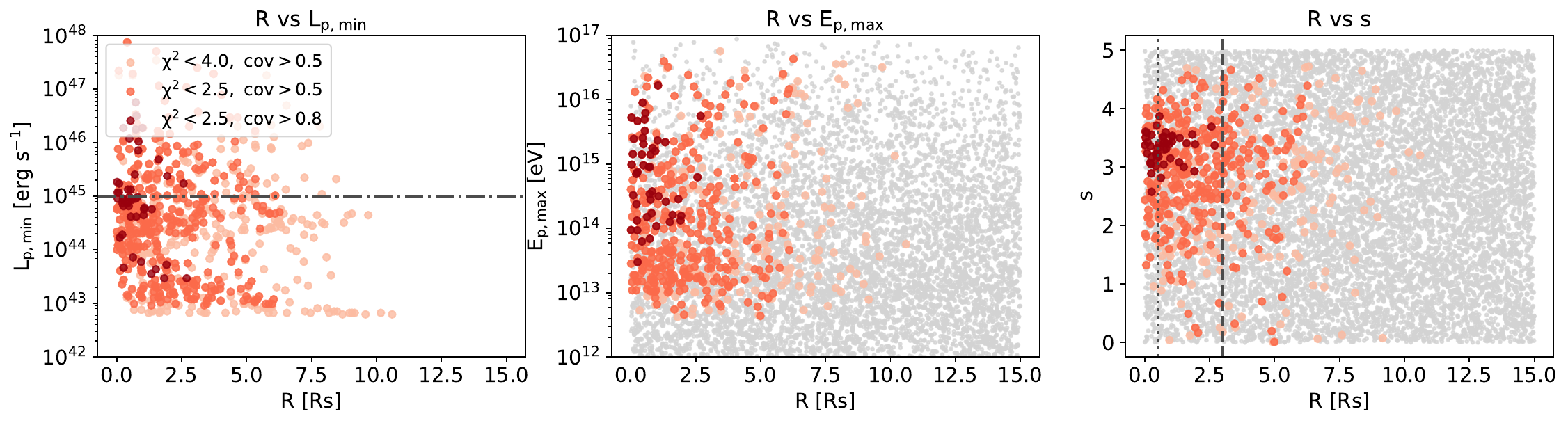}
    \includegraphics[width=\textwidth]{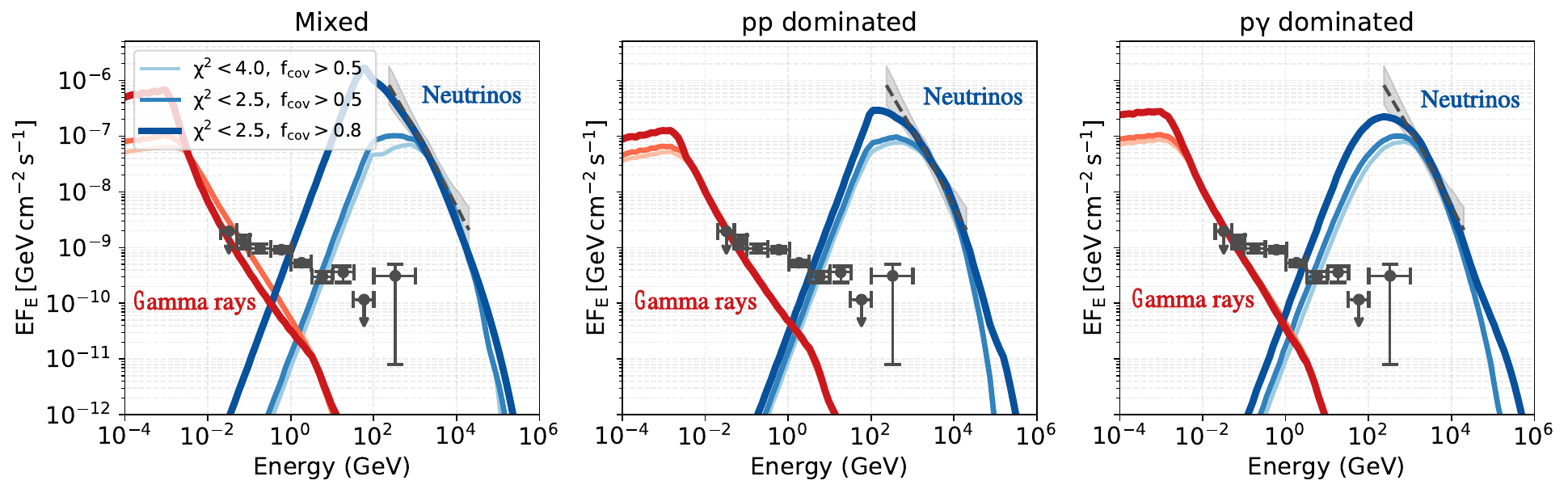}
    
\caption{%
Upper panels: Parameter-space projections of surviving samples for different selection criteria; grey points denote the full LHS ensemble. The horizontal dash-dotted line indicates the bolometric luminosity, while vertical dotted and dashed lines mark $R=0.5R_{\rm s}$ and $R=3R_{\rm s}$, respectively.
Lower panels: Median spectral energy distributions (SEDs) of the corresponding survivors, overlaid with IceCube neutrino data (power-law fit) and Fermi-LAT gamma-ray measurements~\cite{abbasi2026evidence,ajello2023disentangling,das2024revealing}.
Models are labeled \emph{dominated} if a single hadronic channel contributes $>70\%$ of the neutrino flux, and \emph{mixed} otherwise.
}
    \label{fig:sed}
\end{figure*}
\section{Physical setup} 
We adopt a minimal and mechanism-independent framework. Nonthermal protons are assumed to be continuously injected at radius $R$ with an exponential-cutoff-power-law spectrum, $Q_{\rm p}(E_{\rm p}) \propto E_{\rm p}^{-s}\exp(-E_{\rm p}/E_{\rm p,max})$. The injection is normalized to the proton luminosity $L_{\rm p}$,
\begin{equation}
    L_{\rm p} = \int_{E_{\rm p,min}}^{E_{\rm p,max}} E_{\rm p}\, Q_{\rm p}(E_{\rm p})\, dE_{\rm p},
\end{equation}
where the minimum and maximum energies, $E_{\rm p,min}$ and $E_{\rm p,max}$, are treated as free parameters, allowing this parametrization to capture a broad class of proton spectral shapes.

Protons interact with ambient photon fields via $p\gamma$ processes, including X-rays from the corona and optical–infrared emission from the accretion disk. These emissions are modeled following Ref.~\cite{murase2020hidden} and normalized to the observed X-ray luminosity $L_{\rm X} \simeq 5\times10^{43}~\rm erg~s^{-1}$ in the 2--10~keV band and the bolometric luminosity $L_{\rm bol} \simeq 10^{45}~\rm erg~s^{-1}$~\cite{marinucci2015nustar,woo2002active}. In addition, a target gas component with number density $n_0$ is assumed at $R$, associated with coronal plasma or AGN-driven outflows~\cite{inoue2022high,fiorillo2024tev}, leading to $pp$ interactions. We do not assume a priori dominance of either hadronic channel but include both processes self-consistently.

In the steady state, the proton energy distribution can be approximated as $dN_{\rm p}/dE_{\rm p} \simeq Q_{\rm p}(E_{\rm p})\,t_{\rm eff}(E_{\rm p}),$ where the effective timescale is defined by
\begin{equation}
t_{\rm eff}^{-1}
=t_{pp}^{-1}+t_{p\gamma}^{-1}+t_{\rm BH}^{-1}
+t_{\rm syn}^{-1}+t_{\rm esc}^{-1}.
\end{equation}
Here, \(t_{pp}\), \(t_{p\gamma}\), \(t_{\rm BH}\), and \(t_{\rm syn}\) represent the energy-loss timescales associated with \(pp\) interactions, photomeson (\(p\gamma\)) interactions, Bethe--Heitler (BH) pair production, and synchrotron radiation, respectively, while \(t_{\rm esc}\) characterizes the proton escape timescale. For \(pp\) interactions, we adopt
\begin{equation}
t_{pp}=\frac{1}{0.5\sigma_{\rm pp}n_0c},
\end{equation}
where \(\sigma_{\rm pp}\sim5\times10^{-26}\,{\rm cm}^{2}\) is the \(pp\) interaction cross section~\citep{Eidelman2004RPP}, and \(n_0\) is the number density of cold protons. The factor of \(0.5\) accounts approximately for the inelasticity of the \(pp\) interaction. The energy-loss timescale for photomeson production is calculated as
\begin{equation} t^{-1}_{p\gamma} = \frac{c}{2\gamma_p^2} \int_{\tilde{E}_{\rm th}}^{\infty} d\tilde{E} \, \sigma_{p\gamma}(\tilde{E}) \kappa_{p\gamma}(\tilde{E}) \tilde{E} \int_{\tilde{E}/2\gamma_p}^{\infty} dE_{\gamma} \, E_{\gamma}^{-2} \frac{dN_{\gamma}}{dE_{\gamma}}, \end{equation}
where \(\sigma_{p\gamma}\) and \(\kappa_{p\gamma}\) are the photomeson production cross section and inelasticity, respectively~\citep{stecker1968effect,patrignani2016review}, \(\tilde{E}\) is the photon energy in the proton rest frame, and \(\tilde{E}_{\rm th}\simeq145~{\rm MeV}\) is the threshold energy. Here, $\gamma_p = E_p / (m_pc^2) $
is the proton Lorentz factor, and \(dN_{\gamma}/dE_{\gamma}\) denotes the differential number density of the ambient soft photon field. The BH pair-production timescale, \(t_{\rm BH}\), is evaluated using an analogous expression, with the corresponding BH cross section and inelasticity, \(\sigma_{\rm BH}\) and \(\kappa_{\rm BH}\), respectively~\cite{chodorowski1992reaction}. The proton synchrotron cooling timescale is given by
\begin{equation}
t_{\rm syn}
=
\frac{6\pi m_{\rm p}^{4}c^{3}}
{\sigma_{\rm T}m_{\rm e}^{2}B^{2}E_{\rm p}},
\end{equation}
where \(\sigma_{\rm T}\) is the Thomson cross section and \(B\) is the magnetic-field strength in the neutrino-production region. Finally, we parameterize the proton escape timescale as
\begin{equation}
t_{\rm esc}=\eta_{\rm esc}\frac{R}{c},
\end{equation}
with \(\eta_{\rm esc}=10\). We find that our results are insensitive to the adopted value of \(\eta_{\rm esc}\); see Appendix~\ref{sec:esc} for further discussion.

Once the steady-state proton distribution, \(dN_{\rm p}/dE_{\rm p}\), is obtained, we can calculate the production of secondary gamma rays, electrons (positrons), and neutrinos through \(pp\) and \(p\gamma\) interactions, following Ref.~\cite{kelner2006energy} and Ref.~\cite{kelner2008energy}, respectively. Specifically, the neutrino production rate through \(pp\) interactions is given by
\begin{equation}
\label{eq:pp}
\frac{dN_{\nu}}{dE_{\nu}}
=
cn_0
\int_{E_{\nu}}^{\infty}
\sigma_{\rm inel}(E_{\rm p})
\frac{dN_{\rm p}}{dE_{\rm p}}
\Phi_{\nu, pp}\left(\frac{E_{\nu}}{E_{\rm p}},E_{\rm p}\right)
\frac{dE_{\rm p}}{E_{\rm p}},
\end{equation}
where \(\sigma_{\rm inel}\) is the inelastic \(pp\) cross section, and \(\Phi_{\nu, pp}\) is the corresponding neutrino production spectrum. Their definitions and parameterizations are given in Ref.~\cite{kelner2006energy}. For \(p\gamma\) interactions, the neutrino production rate can be written as
\begin{equation}
\label{eq:pg}
\frac{dN_{\nu}}{dE_{\nu}}
=
\int
\frac{dN_{\rm p}}{dE_{\rm p}}
\frac{dN_{\gamma}}{dE_{\gamma}}
\Phi_{\nu,p\gamma}\left(\eta,\frac{E_{\nu}}{E_{\rm p}}\right)
\frac{dE_{\rm p}}{E_{\rm p}}
dE_{\gamma},
\end{equation}
where \(\Phi_{\nu,p\gamma}\) is the neutrino production function for \(p\gamma\) interactions, as defined in Ref.~\cite{kelner2008energy}. The production rates of the other secondary particles can be calculated in an analogous manner using Eq~(\ref{eq:pp}) and Eq~(\ref{eq:pg}).

Hadronic interactions produce gamma-rays and electrons (positrons) that initiate electromagnetic cascades. The steady electron (positron) distribution is governed by the stationary Fokker--Planck equation~\cite{bottcher2013leptonic},
\begin{equation}
\label{eq:fk}
    \frac{\partial}{\partial E_{\rm e}}\left(\dot{E}_{\rm e} N_{\rm e}(E_{\rm e})\right)
    = Q_{\rm e}(E_{\rm e}) + \dot{N}_{\rm e}^{\gamma\gamma}(E_{\rm e}) - \frac{N_{\rm e}(E_{\rm e})}{t_{\rm esc}},
\end{equation}
where $\dot{N}_{\rm e}^{\gamma\gamma}(E_{\rm e})$ denotes pair injection via $\gamma\gamma$ absorption.  The resulting secondary gamma rays and electrons (positrons) from Eq~(\ref{eq:pp}) and Eq~(\ref{eq:pg}) are incorporated into Eq~(\ref{eq:fk}) to solve for the steady-state electron distribution~\cite{bottcher2013leptonic,wang2018hadronic,ma2025seyfert}. The cascade emission is then obtained from synchrotron radiation and inverse Compton scattering of this electron population. The final neutrino and gamma-ray fluxes are denoted by \(F_{\nu}\) and \(F_{\gamma}\), respectively.

\section{Observational constraints} 
The observed neutrino and gamma-ray spectra are used to constrain the models. We digitize the neutrino spectrum and the Fermi-LAT gamma-ray spectrum from Refs.~\cite{abbasi2026evidence,das2024revealing}. For gamma rays, we conservatively require that the model flux does not exceed the observed flux. For neutrinos, we quantify the agreement with observation using two complementary criteria. The first is the reduced chi-square,
\begin{equation}
\label{eq:chi2}
    \chi_{\rm red}^2 = \frac{1}{N_{\rm data}} \sum_{i=1}^{N_{\rm data}} 
    \left( \frac{\log F_{\nu}^{(i)} - \log F_{\rm c}^{(i)}}{\sigma_i} \right)^2,
\end{equation}
where $N_{\rm data}=15$ denotes uniformly spaced energy bins. The central flux is defined as $\log F_{\rm c} = (\log F_{\rm low} + \log F_{\rm high})/2$, with uncertainty $\sigma = (\log F_{\rm high} - \log F_{\rm low})/2$, where $F_{\rm high}$ and $F_{\rm low}$ are the boundary fluxes of the neutrino observation (power-law fit).

This definition assumes approximately Gaussian uncertainties in logarithmic flux space. Under this assumption, $\chi^2$ follows a chi-square distribution with $N_{\rm data}$ degrees of freedom. Models with $\chi_{\rm red}^2 \gg 1$ are disfavored, corresponding to low $p$-values under the null hypothesis. In practice, $\chi_{\rm red}^2 \gtrsim 2.5$ and $\chi_{\rm red}^2 \gtrsim 4.0$ roughly correspond to $\sim3\sigma$ and $\sim5\sigma$ deviations, respectively.

The second criterion is the coverage fraction,
\begin{equation}
\label{eq:fcov}
    f_{\rm cov} = \frac{1}{N_{\rm data}} \sum_{i=1}^{N_{\rm data}} 
    \mathbb{I}\!\left(F_{\nu}^{(i)} \in [F_{\rm low}^{(i)}, F_{\rm high}^{(i)}]\right),
\end{equation}
which measures the fraction of energy bins for which the model prediction falls within the observational band. Physically, $f_{\rm cov}$ directly quantifies the fraction of points contained within the shaded region of the data. It therefore captures the global agreement with the spectral envelope, independent of the exact residuals. Statistically, if the quoted intervals correspond to $\sim1\sigma$ uncertainties and are approximately independent, one expects $f_{\rm cov} \simeq 0.68$ for a consistent model. Significant deviations from this expectation indicate systematic mismatch: $f_{\rm cov} \lesssim 0.5$ suggests strong tension with the data, while $f_{\rm cov} \gtrsim 0.8$ implies that the majority of the model points lie within the observation band.

Together, these criteria provide a robust and model-independent characterization of spectral agreement, capturing both point-wise deviations and global envelope consistency.

\begin{figure*}[t]  
    \centering
    \includegraphics[width=0.8\linewidth,height=0.35\linewidth]{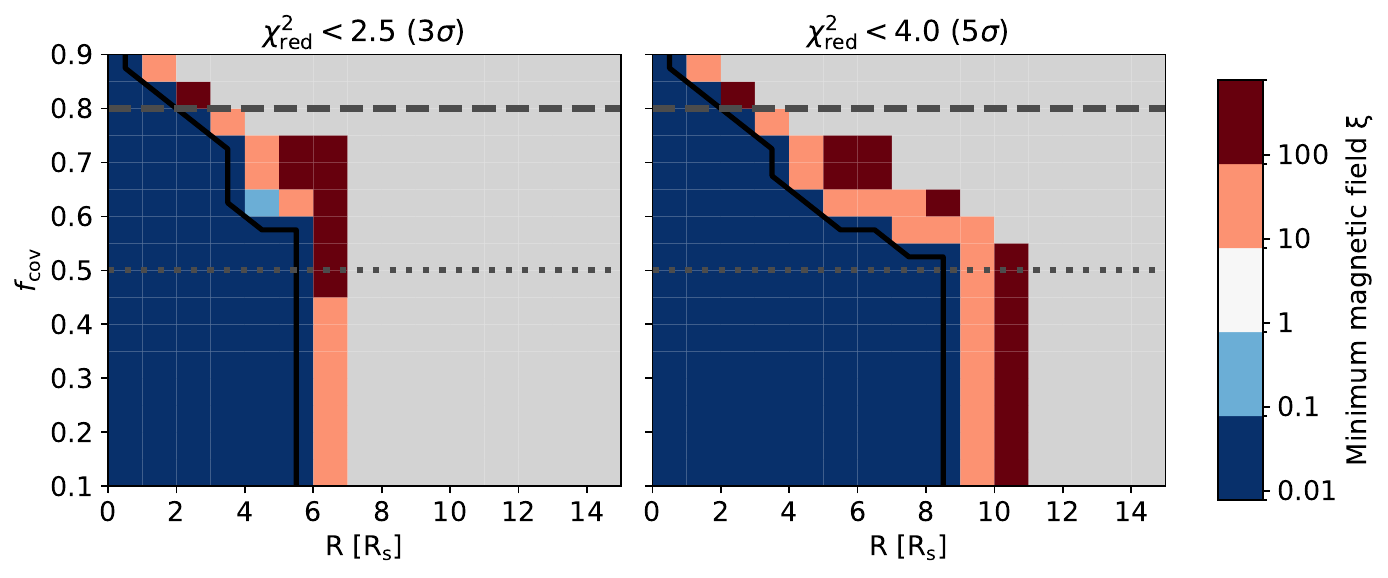}
    \includegraphics[width=0.8\linewidth,height=0.35\linewidth]{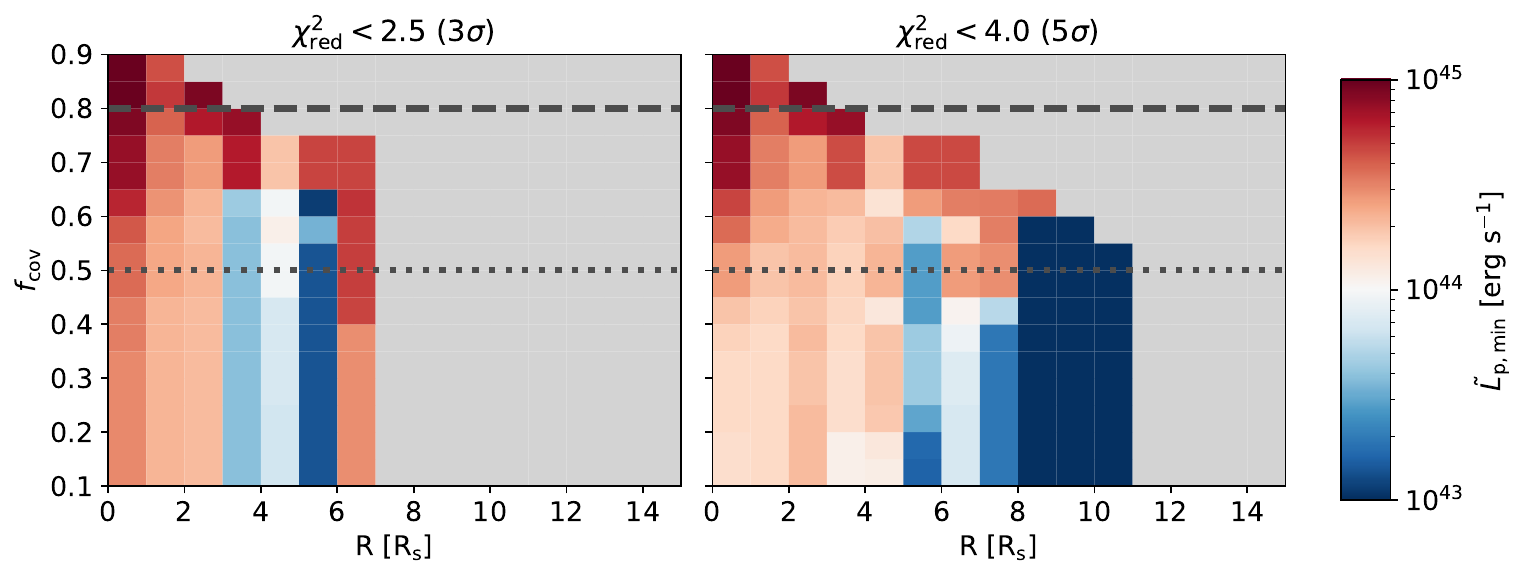}
    \caption{%
Heatmap in the $R$–$f_{\rm cov}$ plane. The emission radius $R$ is binned into 15 intervals (1–15~$R_{\rm s}$), and $f_{\rm cov}$ into 20 bins (0–1). The color scale (blue to red) represents the minimum magnetic field (top panels) or median proton luminosity (bottom panels) required in each bin. Grey regions indicate no surviving points. The black solid line in the upper panels marks approximately the transition between weak- and strong-field regimes.}
    \label{fig:fr}
\end{figure*}

The model flux $F_{\nu(\gamma)}(R,s,E_{\rm p,min},E_{\rm p,max},L_{\rm p},n_0,B)$ depends on seven physical parameters. The magnetic field is parameterized by $\xi \equiv U_B/U_{\rm ph}$~\cite{murase2022hidden}, where $U_{\rm B}$ and $U_{\rm ph}$ are the energy densities for the local magnetic field and photon field, respectively. To efficiently explore this high-dimensional space, we employ Latin hypercube sampling (LHS), which ensures uniform coverage of the parameter ranges listed in Table~\ref{tab:lhs}.

LHS is performed over five parameters, $(R,s,E_{\rm p,min},E_{\rm p,max},n_0)$, with $N_{\rm LHS}=10^4$ samples. For each point, the magnetic parameter is discretely scanned over $\xi =[0.01,0.1,1,10,100]$, yielding $5N_{\rm LHS}$ realizations in total. The convergence for our samplings is verified in Appendix~\ref{sec:LHS}. For each realization, we determine whether a minimum proton luminosity $L_{\rm p,min}$ exists such that the predicted flux satisfies the adopted $\chi_{\rm red}^2$ and $f_{\rm cov}$ criteria.  Models admitting such a condition are retained as \emph{surviving points}. We illustrate the parameter-space projections of surviving points for several criteria in Fig.~\ref{fig:sed}, the corresponding median spectra are also shown.
\begin{table}[t]
\centering
\renewcommand{\arraystretch}{1.2}
\setlength{\tabcolsep}{6pt}  
\caption{Parameter ranges adopted in the Latin hypercube sampling (LHS) and scanning.}
\label{tab:lhs}
\begin{tabular}{lcc}
\toprule
Parameter &notation & Range \\
\midrule
Emission region radius &$R~[R_{\rm s}]$ & $[0,\,15]$ \\
Proton spectral index &$s$ & $[0,\,4]$ \\
Minimum proton energy & $\log E_{\rm p,min}~[\mathrm{eV}]$ & $[9,\,13]$ \\
Maximum proton energy &$\log E_{\rm p,max}~[\mathrm{eV}]$ & $[9,\,18]$ \\
Cold proton density & $\log n_0~[\mathrm{cm^{-3}}]$ & $[0,\,16]$ \\

Minimum  proton luminosity & $\log L_{\rm p,min}~[\mathrm{erg}]$ & [38,50]\\
Magnetic parameter& $\log\xi$ & [-2,2]\\
\bottomrule
\end{tabular}
\end{table}
Across a wide range of criteria ($\chi_{\rm red}^2 = 1$--4, $f_{\rm cov} = 0.1$--0.9), the fraction of surviving points remains small ($\lesssim 7\%$), indicating that the allowed parameter space is highly constrained. For each surviving point, we classify the dominant hadronic process: a channel ($pp$ or $p\gamma$) contributing $>70\%$ of the neutrino flux is labeled \emph{dominated}, otherwise \emph{mixed}. The results (see Appendix~\ref{sec:had}) show that $p\gamma$ interactions are slightly favored over $pp$, while mixed cases are rare.

We then examine the physical conditions of the surviving points. As shown in Fig.~\ref{fig:fr}, the emission region is tightly constrained: $R \gtrsim 10 R_{\rm s}$ is strongly disfavored ($\sim5\sigma$), and $R \gtrsim 6 R_{\rm s}$ is disfavored at the $\sim3\sigma$ level. The required magnetic field increases toward the exclusion boundary, with $\xi > 1$ for large $R$. The emission radius dominates the allowed parameter space, while other parameters yield only subleading constraints. The median minimum proton luminosity is typical $\tilde{L}_{\rm p,min} \sim 10^{43}$--$10^{45}~{\rm erg~s^{-1}}$, and $E_{\rm p,min} \lesssim 10^{12}~{\rm eV}$ is excluded at the $\sim5\sigma$ level.
\begin{table*}[htbp]
\centering
\renewcommand{\arraystretch}{1.4}
\begin{threeparttable}
\caption{%
Median parameter values in the favored region of the parameter space ($f_{\rm cov} > 0.8$ and $\chi^2_{\mathrm{red}} \lesssim 2.5$) for different magnetic-field strengths $\xi$. Quoted uncertainties correspond to the 16th–84th percentiles.}
\label{tab:favor}

\begin{tabular}{lcccccc}
\toprule
$\xi$ & $R~[R_{\rm s}]$ & $\log \tilde{E}_{\rm p,min}~[\mathrm{eV}]$ & $\log \tilde{E}_{\rm p,max}~[\mathrm{eV}]$ & $\log \tilde{L}_{\rm p,min}~[\mathrm{erg~s^{-1}}]$ & $\log \tilde{n_0}$ & $\tilde{s}$ \\
\midrule
0.01 & 0.470$^{+0.571}_{-0.259}$ & 12.025$^{+0.275}_{-0.266}$ & 15.071$^{+0.413}_{-0.828}$ & 44.818$^{+0.226}_{-1.011}$ & 6.256$^{+7.535}_{-4.801}$ & 3.311$^{+0.220}_{-0.254}$ \\
0.1  & 0.482$^{+0.556}_{-0.271}$ & 12.025$^{+0.274}_{-0.381}$ & 14.993$^{+0.474}_{-0.842}$ & 44.818$^{+0.296}_{-0.998}$ & 4.658$^{+9.039}_{-3.189}$ & 3.296$^{+0.233}_{-0.250}$ \\
1    & 0.537$^{+0.453}_{-0.313}$ & 12.019$^{+0.279}_{-0.654}$ & 14.957$^{+0.447}_{-0.658}$ & 44.902$^{+1.048}_{-0.633}$ & 6.519$^{+6.387}_{-4.999}$ & 3.391$^{+0.139}_{-0.313}$ \\
\bottomrule
\end{tabular}

\end{threeparttable}
\end{table*}

\section{Hints for particle acceleration}
As the requirement on $f_{\rm cov}$ is tightened, the allowed parameter space rapidly shrinks, reflecting the increasing difficulty of reproducing the broadband spectrum. We define a \emph{favored} region by $f_{\rm cov} > 0.8$ and $\chi_{\rm red}^2 \lesssim 2.5$, corresponding to models consistent with the data at the $\sim 3\sigma$ level while accounting for $\gtrsim 80\%$ of the observed energy bins.

Within this region, we construct the probability density functions of the model parameters and adopt their median values as representative. The results are summarized in Table~2. The inferred distributions show little dependence on the magnetic parameter $\xi$. The emission region is constrained to $R \lesssim 3 R_{\rm s}$, with typical values approaching the innermost stable circular orbit ($\sim 0.5 R_{\rm s}$). The required proton luminosity reaches $\tilde{L}_{\rm p,min} \sim 10^{45}~{\rm erg~s^{-1}}$, comparable to the bolometric luminosity, while the maximum proton energy is $\tilde{E}_{\rm p,max} \sim 1~{\rm PeV}$. These extreme conditions provide direct constraints on viable acceleration mechanisms.

To assess the feasibility of different acceleration scenarios under these constraints, we impose two physically motivated conditions at $\tilde{E}_{\rm p,max}$: (i) the acceleration timescale equals the effective cooling timescale, $t_{\rm acc} \simeq t_{\rm eff}$, and (ii) the available magnetic power matches the required proton luminosity, $L_B \simeq \tilde{L}_{\rm p,min}$. 

For stochastic acceleration, the acceleration timescales should equal to $t_{\rm eff}$ at $\tilde{E}_{\rm p,max}$, i.e. ,
\begin{equation}
\label{eq:tsa}
    t_{\rm acc,SA}(\eta,B)\simeq\frac{10\eta R}{\sigma_{\rm tur}c}\simeq t_{\rm eff}(\tilde{E}_{\rm p,max}),
\end{equation}
where dimensionless parameter $\eta=l/R$ measures the turbulent coherence length $l$~\cite{fiorillo2024magnetized}. The turbulent magnetization is estimated to be eqaul with magnetization, $\sigma=\sigma_{\rm tur}\simeq B^2/(4\pi n_{0}m_{\rm p}c^2)$, where plasma proton density $n_0$ is equal to electron density $n_{\rm e}$, where $n_{\rm e}\simeq2.5\times10^{10}10R_{\rm s}/(M_7R)~\rm cm^{-3}$ is inferred by the Compton opacity of the corona ~\cite{stern1995geometry,zdziarski1985power,ricci2018bat}.

In the optimistic assumption, all the turbulent magnetic
energy dissipates to nonthermal protons, yielding (e.g., Ref.~\cite{fiorillo2024magnetized}) 
\begin{equation}
\label{eq:lsa}
    L_{\rm B,SA}(\eta,B)\simeq\frac{2\pi}{3}\frac{0.1}{\eta}\frac{c^3\sigma^{3/2}_{\rm tur}}{(1+\sigma_{\rm tur})^{1/2}}n_0m_{\rm p}R^2\simeq \tilde{L}_{\rm p,min}.
\end{equation}
Similarly, for the magnetic reconnection scenario, we have
\begin{equation}
\label{eq:tmr}
    t_{\rm acc,MR}(\beta_{\rm rec},B)\simeq\frac{\tilde{E}_{\rm p,max}}{qBc\beta_{\rm rec}}\simeq t_{\rm eff}(\tilde{E}_{\rm p,max}),
\end{equation}
and 
\begin{equation}
\label{eq:lmr}
    L_{\rm B,MR}(\beta_{\rm rec},B)\simeq \beta_{\rm rec}\frac{B^2}{8\pi}cR^2\simeq \tilde{L}_{\rm p,min},
\end{equation}
where $\beta_{\rm rec}$ represents the plasma inflow
speed into the reconnection layer, normalized to the speed of
light.

For SA, combining Eqs.~(\ref{eq:tsa}) and (\ref{eq:lsa}) yields physically reasonable solutions within the favored region. For example, at $R = R_{\rm s}$, typical parameters such as $\eta \sim 0.09$ and $B \sim 10^5~{\rm G}$, satisfy both the timescale and energetics constraints. In contrast, the MR scenario using Eqs.~(\ref{eq:tmr}) and (\ref{eq:lmr}) leads to solutions that require extreme parameters, with $B \sim 10^{9}$--$10^{10}~{\rm G}$ and $\beta_{\rm rec} \sim 10^{-10}$--$10^{-9}$, far below the commonly expected value $\beta_{\rm rec} \sim 0.1$~\cite{comisso2016value,cassak2017review}. Such values are difficult to reconcile with current theoretical and simulation results.

This comparison suggests that, within the compact and high-luminosity conditions implied by the favored parameter space, stochastic acceleration can more naturally account for the required physical conditions, whereas magnetic reconnection appears disfavored due to its requirement for extreme parameter values.

We emphasize that our result favoring  SA does not imply that magnetic reconnection is ruled out. Our conclusion is based on relatively stringent constraints on the neutrino spectral shape: the model spectrum is required to closely follow the IceCube power-law confidence band across the full energy range, using the latest data extending down to \(\sim0.2\) TeV. Under these constraints, the allowed parameter space favors SA, for which the required conditions are reasonably close to typical values. In contrast, relaxing the spectral constraints may allow magnetic reconnection to provide a viable explanation, or at least a partial contribution, as demonstrated by Refs.~\cite{das2026multimessenger,passos2026turbulence}. The differences from previous studies therefore mainly arise from the adopted spectral constraints and IceCube data. In particular, previous works may place less emphasis on reproducing the detailed power-law spectral shape or use earlier data with a narrower energy range. Our results should thus be interpreted as indicating that SA provides a more natural explanation under the updated and stringent spectral constraints, rather than excluding magnetic reconnection as a possible neutrino production mechanism.

DSA is not considered here, as $L_{\rm p}$ is not directly constrained by the local magnetic energy budget. Moreover, the compact emission region ($R \lesssim 3 R_{\rm s}$) disfavors the formation of shocks by outflows~\cite{costa2020powering,miller2016accretion}. Although a jet is suggested in this source, its kinetic power ($\sim 10^{43}~{\rm erg~s^{-1}}$) falls well below the required $\tilde{L}_{\rm p,\rm min} \sim 10^{45}~{\rm erg~s^{-1}}$~\cite{salvatore2024possible}. Accretion shocks may exist, but are not commonly seen in current simulations of inner accretion flows~\cite{gu2006non,garain2025general}. These considerations further limit the viability of DSA.
\section{Summary and discussion}
We adopt a minimally model-independent framework to infer the physical conditions of neutrino production in NGC~1068, primarily constrained by the observed spectral shape. The emission region is tightly limited: $R \gtrsim 10 R_{\rm s}$ is strongly disfavored ($\sim5\sigma$), and $R \gtrsim 6 R_{\rm s}$ is excluded at the level $\sim3\sigma$. The required proton luminosity is typically $\sim0.01-0.1 L_{\rm bol}$.

Imposing a more stringent requirement ($\gtrsim80\%$ of the neutrino spectrum be reproduced) further tightens the constraints to $R \lesssim 3 R_{\rm s}$, with $L_{\rm p} \sim L_{\rm bol}$ and $E_{\rm p,max} \sim 1~{\rm PeV}$. Under these extreme conditions, stochastic acceleration remains viable, while alternative mechanisms require implausible parameters. These results indicate that stochastic acceleration is the most natural mechanism for producing the observed high-energy neutrinos in this source.

Our analysis differs from previous studies in three main respects. First, compared with Ref.~\cite{carpio2026multimessenger,testagrossa2026neutrinos}, whose best-fit spectra can deviate noticeably from the power-law SED reported by IceCube, our criteria, defined by Eqs.~(\ref{eq:chi2}) and (\ref{eq:fcov}), place greater emphasis on the spectral shape. We require the model spectrum to remain consistent with the IceCube confidence band over the full energy range, rather than simply selecting the statistically preferred spectrum. Second, unlike Ref.~\cite{das2024revealing,das2026multimessenger}, which fix several parameters to make direct parameter-space scans computationally feasible, we employ Latin hypercube sampling to efficiently explore the full seven-dimensional parameter space without fixing parameters a priori. Finally, we adopt the latest neutrino spectral measurements from Ref.~\cite{abbasi2025icecube}, which extend from \(\sim0.2\) to \(\sim20\) TeV while maintaining a nearly power-law shape. This broader energy coverage, together with our stricter spectral criterion, provides a stronger constraint on the model spectrum and consequently leads to a more tightly constrained parameter space. In particular, these differences can account for our more compact emission-region constraint of \(R\lesssim3R_S\), compared with the preferred \(R\sim7.5R_S\) found by Ref.~\cite{das2026multimessenger}.

Importantly, our conclusion is a statistical inference based on current observations and can be refined with future data from next-generation neutrino detectors such as IceCube-Gen2~\cite{aartsen2021icecube} and KM3NeT~\cite{margiotta2014km3net}, as well as upcoming MeV–sub-GeV gamma-ray missions including e-ASTROGAM~\cite{de2018science} and AMEGO~\cite{fleischhack2021amego}.

\begin{acknowledgments}
We thank the anonymous reviewer for his helpful comments and suggestions. We acknowledge support from the National Natural Science Foundation of China under grant No.12003007 and the Fundamental Research Funds for the Central Universities (No. 2020kfyXJJS039).
\end{acknowledgments}

\begin{figure}[htbp]
\centering
\includegraphics[width=1.0\linewidth,height=0.8\linewidth]{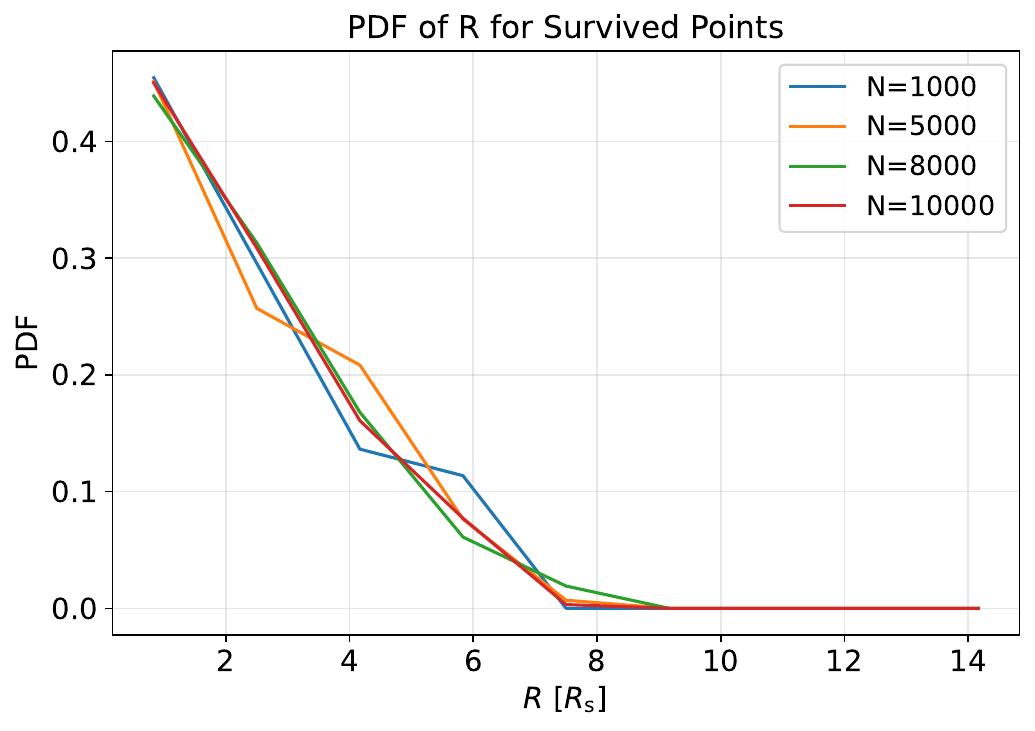}
\caption{Probability density of the emission radius $R$ for surviving samples satisfying $\chi_{\rm red}^2 < 2.5$ and $f_{\rm cov} > 0.5$, shown for increasing LHS sample sizes $N_{\rm LHS}=10^3$, $5\times10^3$, $8\times10^3$, and $10^4$. The convergence of the distribution at large $N_{\rm LHS}$ indicates sufficient sampling of the parameter space.}
\label{fig:cov}
\end{figure}

\appendix
\begin{figure*}[htbp]
\centering
\includegraphics[width=1.0\linewidth,height=0.3\linewidth]{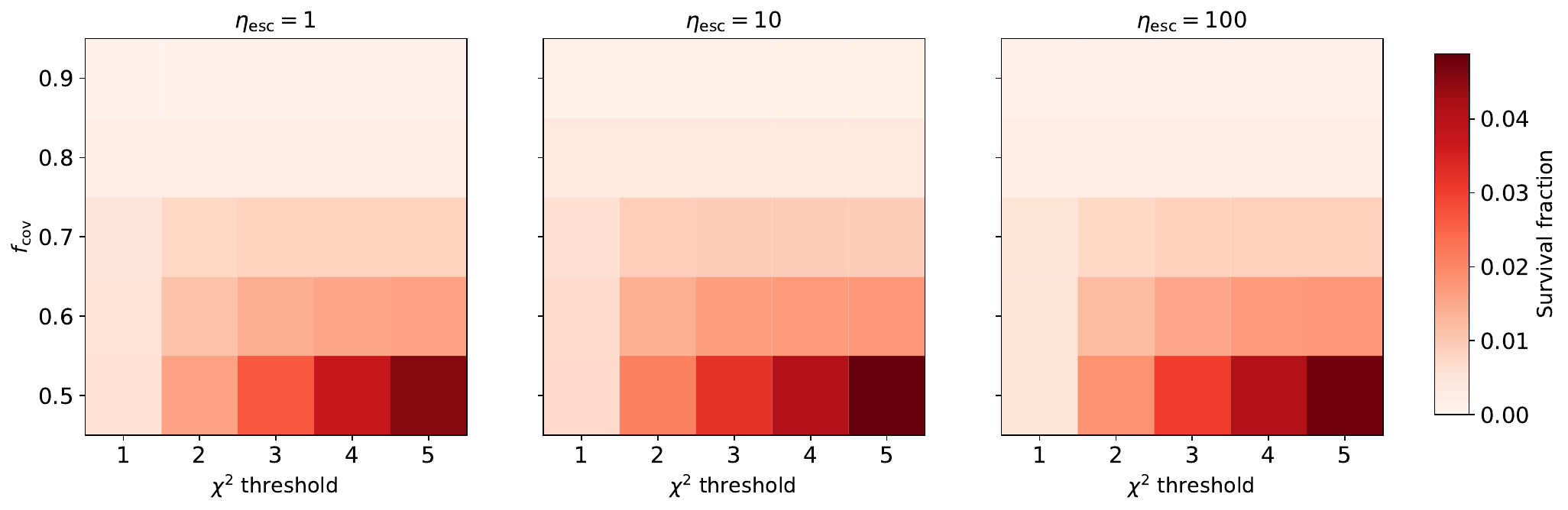}
\caption{Surviving fraction $N_{\rm sur}/N_{\rm LHS}$ under combined $\chi_{\rm red}$ and $f_{\rm cov}$ criteria for different escape efficiencies $\eta_{\rm esc}=1,\,10,$ and $100$. The distributions show no significant dependence on $\eta_{\rm esc}$, indicating a negligible impact of particle escape on the allowed parameter space.}
\label{fig:eta}
\end{figure*}

\begin{figure*}[htbp]
\centering
\includegraphics[width=0.8\linewidth,height=0.6\linewidth]{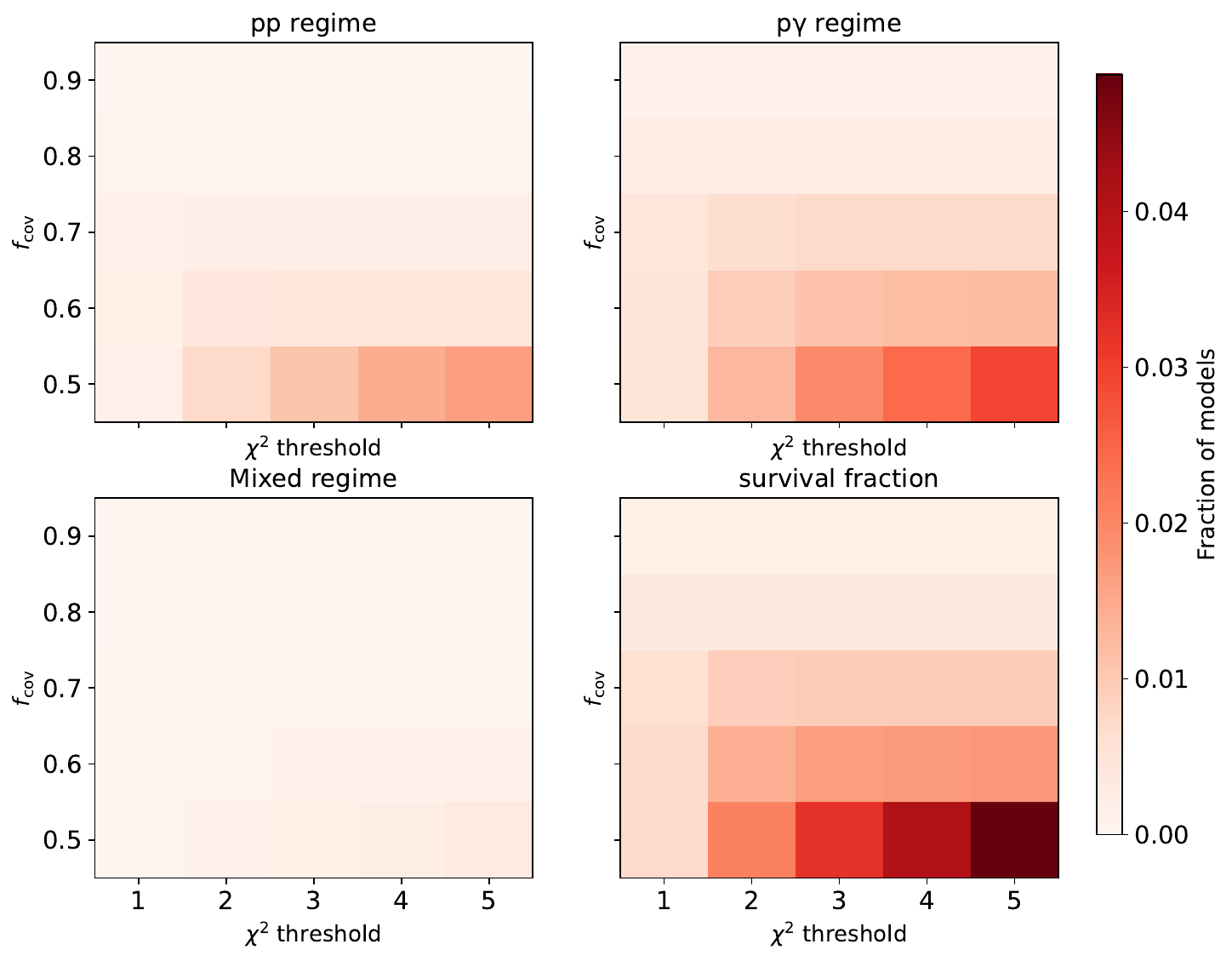}
\caption{Surviving fraction $N_{\rm sur}/N_{\rm LHS}$ under combined $\chi_{\rm red}$ and $f_{\rm cov}$ criteria for each dominant channel ($pp$, $p\gamma$, and mixed), together with the total. Mixed-channel solutions are strongly suppressed, while single-channel scenarios dominate, with a moderately larger fraction for $p\gamma$.}
\label{fig:dom}
\end{figure*}
\section{Particle Escape}
\label{sec:esc}
For charged particles (protons, electrons, and positrons), we parametrize the escape timescale as $t_{\rm esc}=\eta_{\rm esc} R/c$, where $\eta_{\rm esc}$ characterizes the escape efficiency. We adopt $\eta_{\rm esc}=10$ as the fiducial value in the main text, corresponding to a moderately inefficient escape.

To assess the robustness of our results against this assumption, we explore a wide range of escape efficiencies, $\eta_{\rm esc}=1,\,10,$ and $100$, spanning the regimes from efficient to inefficient escape. Figure~\ref{fig:eta} shows the surviving fraction $N_{\rm sur}/N_{\rm LHS}$ under different combined $\chi_{\rm red}$ and $f_{\rm cov}$ criteria for these values.

The distributions remain essentially unchanged across this range of $\eta_{\rm esc}$, demonstrating that particle escape has a negligible impact on the selection of viable parameter space. This indicates that the constraints derived in this work are primarily driven by radiative and interaction processes rather than escape effects.
\section{Latin Hypercube Sampling}
\label{sec:LHS}

To efficiently explore the high-dimensional parameter space, we adopt Latin hypercube sampling (LHS), a stratified sampling technique designed to ensure uniform coverage of each parameter range. In LHS, the domain of each parameter is partitioned into equal-probability intervals, and samples are drawn such that each interval is represented exactly once along each dimension. This construction suppresses clustering and yields a more uniform coverage of the multidimensional space compared to purely random sampling.

LHS is particularly well-suited for problems with moderate dimensionality. In the present work, the parameter space is five-dimensional, for which the required sample size must be explicitly validated to ensure convergence. To this end, we examine the probability density distribution of the emission radius $R$ for surviving points under the selection criteria $\chi_{\rm red}^2 < 2.5$ and $f_{\rm cov} > 0.5$, for increasing sample sizes $N_{\rm LHS} = 10^3,\,5\times10^3,\,8\times10^3,$ and $10^4$ (Fig.~\ref{fig:cov}).

As shown in Fig.~\ref{fig:cov}, the distributions obtained with $N_{\rm LHS}=8\times10^3$ and $ N _ {\rm LHS} = 10^4$ are nearly indistinguishable, indicating that the sampling has converged at this level. In contrast, smaller sample sizes exhibit noticeable fluctuations. We therefore adopt $N_{\rm LHS}=10^4$ in the main analysis, which provides sufficient coverage of the parameter space while remaining computationally tractable.

\section{Hadronic Channels}
\label{sec:had}
For each surviving point, we determine the dominant neutrino production channel by integrating the $pp$ and $p\gamma$ neutrino fluxes over the energy range $[0.2,\,20]$~TeV. This energy range is chosen to match the observational energy band that primarily constrains the spectral shape.

We define a channel as \emph{dominated} if its integrated contribution exceeds $70\%$ of the total neutrino flux, i.e., $F_i / (F_{pp}+F_{p\gamma}) > 0.7$, where $F_i$ denotes the integrated flux of channel $i$. Otherwise, the emission is classified as \emph{mixed}. This criterion provides a quantitative and model-independent way to identify the leading production mechanism.

In Fig.~\ref{fig:dom}, we show the surviving fractions under combined $\chi_{\rm red}$ and $f_{\rm cov}$ criteria for each dominant channel, together with the total. Mixed-channel solutions are strongly suppressed, indicating that single-channel scenarios more readily reproduce the observed power-law spectrum. We further find that the $p\gamma$ channel exhibits a moderately higher surviving fraction than the $pp$ channel. This indicates that, under current observational constraints, the mechanism $p\gamma$ occupies a larger viable region of the parameter space. However, $pp$-dominated solutions remain present and cannot be excluded.

\bibliography{reference}

\end{document}